\documentclass{article}
\makeatletter
\@addtoreset{equation}{section}

\makeatother

\usepackage{graphics}

\begin{document}

\begin{flushright}
Aug 2007

KUNS-2089
\end{flushright}

\begin{center}

\vspace{3cm}

{\LARGE 
\begin{center}
Some Implications of Perturbative Approach 

to $AdS$/CFT Correspondence
\end{center}
}

\vspace{2cm}

Hikaru Kawai $^{ab}$\footnote{e-mail address : hkawai@gauge.scphys.kyoto-u.ac.jp} \hspace{2mm}
and 
\hspace{2mm}
Takao Suyama $^a$\footnote{e-mail address : suyama@gauge.scphys.kyoto-u.ac.jp}

\vspace{1cm}

$^a$ {\it Department of Physics, Kyoto University,}

{\it Kitashirakawa, Kyoto 606-8502, Japan }

{\it and}

$^b$ {\it Theoretical Physics Laboratory,} 

{\it The Institute of Physics and Chemical Research (RIKEN),}

{\it Wako, Saitama 351-0198, Japan}

\vspace{3cm}

{\bf Abstract} 

\end{center}

We show some implications of the approach to $AdS$/CFT correspondence 
based on Type IIB string in the flat space-time with D3-branes 
proposed in our previous paper. 
We discuss a correspondence for high energy scattering amplitudes of ${\cal N}=4$ super-Yang-Mills 
proposed recently. 
We also discuss $AdS$/CFT correspondence at finite temperature. 
Our approach provides clear understanding of these issues.

\newpage

\vspace{1cm}

\section{Introduction}

\vspace{5mm}

In our previous paper \cite{KS}, we proposed how to understand the reason why $AdS$/CFT correspondence 
holds. 
Our discussion there was based on the perturbative Type IIB string in the flat space-time with 
D3-branes introduced as boundaries of the worldsheets. 
We mainly discussed the relation 
\cite{RY}\cite{Maldacena} between Wilson loops in ${\cal N}=4$ super-Yang-Mills 
(SYM) in four dimensions and minimal surfaces in the $AdS_5\times S^5$ space-time, and showed that 
the correspondence is a consequence of an approximate symmetry which 
exists in the worldsheet theory, 
if the large $N$ limit is taken with the 't Hooft coupling $\lambda$ kept finite but large, and if the 
worldsheets relevant for evaluating the Wilson loops and the minimal surfaces are restricted within 
a region near D3-branes. 

In this paper, we show that our approach is useful for understanding the other issues discussed in the 
literature. 
The first example, discussed in section 2, 
is on the calculation of high energy scattering amplitudes of the SYM, which was proposed in 
\cite{Dixon} and recently confirmed on the strong coupling side at the level of four-gluon scattering 
in \cite{AM}. 
It will be shown that this proposal is also justified by the scale invariance in a similar way to 
the case of the Wilson loop. 
The second  example is on $AdS$/CFT correspondence at finite temperature, discussed in section 3. 
The discussions on Wilson loops can be carried out also in this case. 
From the point of view of the perturbative string with D3-branes, the issue on the entropy can be 
understood rather clearly.

\vspace{1cm}

\section{High energy scattering amplitudes}

\vspace{5mm}

Let us consider a scattering process of $n$ massless open string states 
living on parallel $N$ D3-branes. 
At a low energy scale, the dynamics of open strings is governed by ${\cal N}=4$ SYM in four 
dimensions. 
The scattering amplitude can be calculated from worldsheets with a number of 
boundaries on which $n$ vertex operators for the massless states are attached. 
We only consider the case in which all the vertex operators are attached to a single boundary of the 
worldsheet. 
Let us take the large $N$ limit with the 't Hooft coupling $\lambda=g_sN$ fixed. 
Due to taking this limit, closed string handles of the worldsheets are suppressed, and therefore, 
the relevant topology of the worldsheets is that of the disk with boundaries. 
These worldsheets correspond to planar ribbon graphs of the SYM. 
Let $(t^{a_k})_{ij}$ be the color-dependent factor of the $k$-th vertex operator. 
Then the amplitude considered here 
is proportional to Tr$(t^{a_1}t^{a_2}\cdots t^{a_n})$ which is the only factor 
depending on the gauge group. 

\begin{figure}[tbp]
\hspace*{1.5cm}
\includegraphics{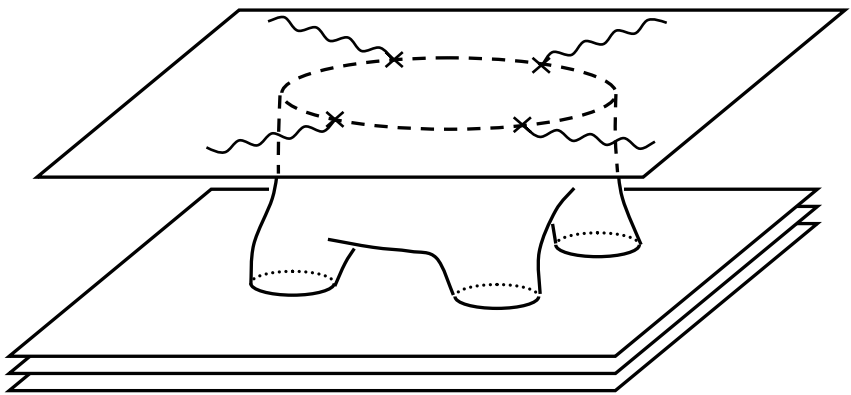}
\caption{Worldsheet configuration for the regularized amplitude. }
   \label{worldsheet}
\end{figure}

\begin{figure}[tbp]
\hspace*{2cm}
\includegraphics{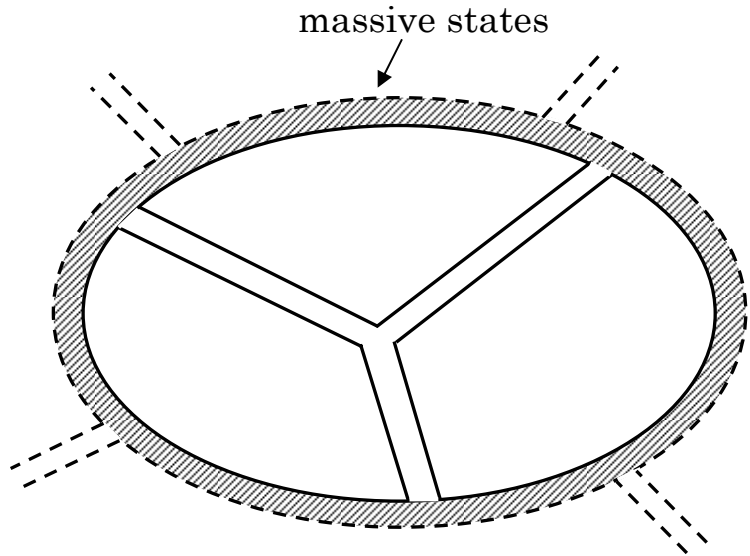}
\caption{Feynman diagram corresponding to the worldsheet in figure \ref{worldsheet}. }
   \label{Feynman}
\end{figure}

Suppose that one of the D3-branes is separated from the other $N-1$ D3-branes which are on top of 
each other, and all of the external open string states are on the separated D3-brane. 
The restriction of the external states only results in choosing a trivial 
prefactor, instead of a generic Tr$(t^{a_1}t^{a_2}\cdots t^{a_n})$, 
and therefore, it is easy to deduce the generic amplitude from this special one. 
We can also restrict the sum over worldsheets so that the boundaries, on which there is no vertex 
operator, are on the coincident D3-branes. 
This restriction results in assigning $N-1$, not $N$, to each color index 
loop, and therefore, this leads 
to a modification of the amplitude by an amount of order $\frac1N$, which is negligible in the 
large $N$ limit. 
Due to the separation of the D3-brane, the strips of the string corresponding to the outermost loop 
are stretched between the D3-branes with a finite width, and therefore, it is a massive state that 
is propagating along this loop. 
The situation is depicted in figure \ref{worldsheet}. 
The figure \ref{Feynman} shows the corresponding Feynman diagram. 

The introduction of mass in this manner is expected to make the amplitude 
well-defined in the IR region. 
In terms of the ${\cal N}=4$ SYM, what we have done is the 
following. 
We first give a nonzero background to one of the scalar fields as
\begin{equation}
\begin{array}{cc}
  & 
  \begin{array}{cccc} 1&\ldots & n & n+1 \ldots N \\ \end{array} 
  \\
  \begin{array}{c}
    1     \\
    \vdots\\
    n     \\
    n+1   \\
    \vdots\\
    N     \\
  \end{array}
  &
  \left[
    \begin{array}{cccccc}
         \mu &      &    &    &      &    \\
             &\ddots&    &    &      &    \\
             &      & \mu&    &      &    \\
             &      &    &  0 &      &    \\
             &      &    &    &\ddots&    \\
             &      &    &    &      & 0  \\
     \end{array}
  \right]
  \\ 
\end{array} ,
\end{equation}
where the situation depicted in figure \ref{worldsheet} corresponds to $n=1$, and one can consider a 
situation for generic $n$. 
Here we assume $n \ll N$, and consider only planar diagrams.
We further assume that the colors of external particles
are taken from the $n\times n$ matrices.
Therefore, in the double line notation, the index of the 
outermost index loop runs from $1$ to $n$ .
Furthermore we can constrain
the other index loops to run from $n+1$ to $N$.
This constraint does not affect the amplitude if  $n \ll N$ .
In this way, we give non-zero mass to all the propagators 
that belong to the outermost loop as in fig.2.

\begin{figure}[tbp]
\hspace*{3cm}
\includegraphics{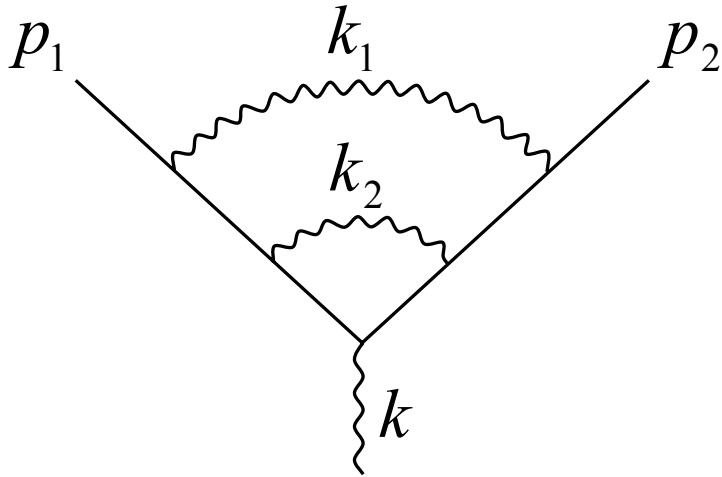}
\caption{A Feynman diagram which could have an IR divergence. }
   \label{IR}
\end{figure}

In order to illustrate the mechanism that mass of the 
outermost propagators regularizes the IR divergence, we consider the following integral which 
corresponds to the Feynman diagram depicted in figure \ref{IR}, 
\begin{equation}
I = \int {d^4 k_1 }\frac{1}
{{k_1 ^2 }}\frac{1}
{{\left( {p_1  - k_1 } \right)^2 }}\frac{1}
{{\left( {p_2  + k_1 } \right)^2 }}
\int {d^4 k_2}
\frac{1}{{k_2 ^2 }}\frac{1}
{{\left( {p_1  - k_1  - k_2 } \right)^2 }}\frac{1}
{{\left( {p_2  + k_1  + k_2 } \right)^2 }} .
\end{equation}
Here we assume $k^2\neq 0$ for simplicity.
If $p_1^2=p_2^2=0$ , the integral of $k_2$ gives 
a term of order $\log(k_1^2)$, 
and the $k_1$ integral is IR divergent.
The regularization we are discussing is 
to give mass to all propagators in the outermost loop.
However, it is sufficient to consider only one propagator in this 
simple case. 
Actually, if we give mass $\mu$
to the outer gluon, $I$ is regularized to
\begin{equation}
I_{reg}= \int {d^4 k_1 }\frac{1}
{{k_1 ^2+\mu^2 }}\frac{1}
{{\left( {p_1  - k_1 } \right)^2 }}\frac{1}
{{\left( {p_2  + k_1 } \right)^2 }}
\int {d^4 k_2}
\frac{1}
{{k_2 ^2 }}\frac{1}
{{\left( {p_1  - k_1  - k_2 } \right)^2 }}\frac{1}
{{\left( {p_2  + k_1  + k_2 } \right)^2 }} .
\end{equation}
The $k_1$ integral converges this time,
although the $k_2$ integral gives $\log(k_1^2)$.
The point is that the mass of an outer propagator makes 
the whole diagram IR finite.\footnote{
As far as we know, there is no proof for the 
justification of the regularization procedure described above. 
It is very interesting to prove that our procedure works for generic amplitudes. 
}. 

Note that the mass introduced for the IR regularization must be much smaller than the string scale, 
and therefore, the separation among D3-branes must be very small. 
Note also that the momenta of the external states are also much smaller than the string scale so as 
not to produce massive string states in the scattering process, although we call this process a 
``high energy'' scattering.

\vspace{5mm}

To relate the scattering amplitude to a classical worldsheet configuration, 
let us make the following change of variables in the worldsheet theory, 
\begin{eqnarray}
\partial_\alpha X^\mu &=& \epsilon_{\alpha\beta}\partial_\beta X_D^\mu, 
   \label{dual} \\
X^I &=& X_D^I, 
   \label{dual2}
\end{eqnarray}
for $\mu=1,\cdots,4$ and $I=6,\cdots,10$. 
This is nothing but the T-duality transformation of the worldsheet variables. 
The transformation of the fermionic variables is defined as 
\begin{eqnarray}
S^a &=& S^a_D, 
   \label{dualS1} \\
\tilde{S}^a &=& M^{ab}\tilde{S}^b_D. 
   \label{dualS2}
\end{eqnarray}
We have employed the Green-Schwarz formalism and taken the light-cone gauge. 
Since we consider worldsheets in the flat space-time, this transformation obviously preserves the 
worldsheet action. 
The boundary condition for $X^\mu$ turns into the Dirichlet boundary condition since (\ref{dual}) 
implies 
\begin{equation}
\partial_\sigma X^\mu = \partial_\tau X^\mu_D, 
   \label{exchange}
\end{equation}
at the boundaries. 
As a result, the D3-branes in the original setup turn into D-instantons. 
It is interesting to notice that the constant modes $x_{D}^\mu$ of $X_D^\mu$ are not fixed, since 
the boundary condition is $\partial_\tau X_D^\mu=0$. 
Therefore, $x_{D}^\mu$ should be integrated in the worldsheet path-integral 
which indicates that the D-instantons are distributed uniformly along $x^\mu$-directions. 

It is very interesting to compare this situation with the large $N$ reduction of the SYM. 
In the reduced model, the momentum integrals for Feynman diagrams correspond to integrals of diagonal 
elements of matrices \cite{GK}. 
If the reduced model is regarded as an effective theory of D-instantons, then the diagonal elements 
of the matrices dictate the positions of the D-instantons. 
This chain of correspondences also implies that the integration of the positions of the D-instantons 
is necessary 
in the calculation of the high energy scattering amplitude. 

In \cite{AM}, classical solutions for the worldsheet are discussed in $AdS_5\times S^5$ background, and 
then the T-dual transformation is performed to obtain explicit expressions of the solutions. 
This corresponds, in our perturbative 
approach, to performing an anisotropic scale transformation of \cite{KS} 
to go to the gravity region, 
and then perform the above transformation. 
In this paper, we proceed through a different way; 
we perform the above T-dual transformation first, and then perform a scale transformation 
defined below. 

It should be noted that the transformation we would like to perform is not exactly the T-duality 
transformation, since the $x^\mu$-directions are non-compact. 
The analysis below is thus valid only when we restrict ourselves with the worldsheets which are 
topologically a disk with boundaries. 
Inclusion of handles, corresponding to considering a finite $N$ case, would require a more complicated 
analysis. 
In the following, however, we call this the T-duality transformation, which would probably make no  
confusion. 

\vspace{5mm}

From now on, we consider disk worldsheets in the presence of $N$ D-instantons. 
As mentioned 
above, this system is equivalent to the original system including D3-branes only in the large $N$ 
limit. 

We consider the scale transformation discussed in \cite{KS} in this D-instanton system. 
The transformation properties of the T-dual variables can be easily derived from the original one. 
As will be shown below, it is an isotropic scale transformation in this case. 
However, this transformation will 
turn out to be an approximate symmetry in a similar sense to \cite{KS}. 
As a result, we will verify that 
the high energy scattering amplitude of ${\cal N}=4$ SYM can be calculated by a specific configuration 
of the worldsheet in $AdS_5$, as is claimed in \cite{AM}. 

Recall that the scale transformation of the coordinate fields in the D3-brane setup is 
\begin{eqnarray}
\delta X^i(\sigma) &=& M^{ij}X^j(\sigma), \\
\delta P^i(\sigma) &=& -M^{ij}P^j(\sigma), \\
\delta S^a(\sigma) &=& iM^{ab}\tilde{S}^b(\sigma), \\
\delta \tilde{S}^a(\sigma) &=& -iM^{ab}S^b(\sigma), 
\end{eqnarray}
where $i,j,a,b$ run from 1 to 8, and the $8\times 8$ matrix $M^{ij}$ and $M^{ab}$ are defined as 
\begin{eqnarray}
M^{ij} &=& \left[ 
\begin{array}{cc}
-I_{4\times 4} & 0 \\ 0 & I_{4\times 4} 
\end{array}
\right], \\
M^{ab} &=& (\gamma^1\gamma^2\gamma^3\gamma^4)^{ab}, 
\end{eqnarray}
where $\gamma^i$ are the $SO(8)$ gamma matrices. 

The T-duality transformation (\ref{dual})(\ref{dual2})(\ref{dualS1})(\ref{dualS2}) provides the scale 
transformation of the T-dual variables as 
\begin{eqnarray}
\delta X^i_D(\sigma) &=& X^i_D(\sigma), \\
\delta P^i_D(\sigma) &=& -P^i_D(\sigma), \\
\delta S^a_D(\sigma) &=& i\tilde{S}^a_D(\sigma), \\
\delta \tilde{S}^a_D(\sigma) &=& -iS^a_D(\sigma). 
\end{eqnarray}
Note that the transformation (\ref{dual}) exchanges the coordinates and the momenta which is clear in 
(\ref{exchange}). 
In this T-dual situation, the scale transformation is indeed the ordinary 
isotropic scale transformation. 
One can easily obtain the scale transformation of the oscillators as 
\begin{eqnarray}
\delta\alpha_n^i &=& -\tilde{\alpha}_{-n}^i, \\
\delta\tilde{\alpha}_n^i &=& -\alpha_{-n}^i, \\
\delta S^a_n &=& i\tilde{S}^a_{-n}, \\
\delta\tilde{S}^a_n &=& -iS^a_{-n}. 
\end{eqnarray}
Using these transformation rules, 
it is easy to show that the boundary state of a D-instanton 
\begin{eqnarray}
|B\rangle &=& \exp\left[ \sum_{n=1}^\infty\left( \frac1n \alpha_{-n}^i\tilde{\alpha}_{-n}^i
 -iS^a_{-n}\tilde{S}^a_{-n} \right) \right]|B_0\rangle, \\
|B_0\rangle &=& \left( |i\rangle|\tilde{i}\rangle-i|\dot{a}\rangle|\tilde{\dot{a}}\rangle \right)
 |x^I=0\rangle|p^\mu=0\rangle\left(\bigotimes_{n=1}^\infty|0_n\rangle \right)
\end{eqnarray}
is invariant under the scale transformation. 
Note that 
\begin{equation}
|p^\mu=0\rangle = \int d^4q\ |x^\mu = q^\mu\rangle
\end{equation}
describes the uniform distribution of the D-instanton. 

\vspace{5mm}

To show the existence of the scale invariance in the D-instanton case, let us consider the 
free energy in the D-instanton background defined as 
\begin{equation}
F(\lambda) = \sum_{n=0}^\infty \frac{F_n}{n!}\lambda^n, 
   \label{free}
\end{equation}
where $F_n$ contains the contributions from worldsheets with $n$ boundaries. 

It is easy to calculate the variation $\delta S$ 
of the worldsheet action under the scale transformation, and 
it can be shown that the variation is a sum of vertex operators corresponding to the state 
$|B_0\rangle$. 
Since the boundary state is invariant, the variation of $F_n$ under the scale transformation 
is obtained by inserting the state $|B_0\rangle$ on the worldsheet, that is, 
\begin{equation}
\delta F_n = F_n(|B_0\rangle). 
\end{equation}

\begin{figure}[tbp]
\includegraphics{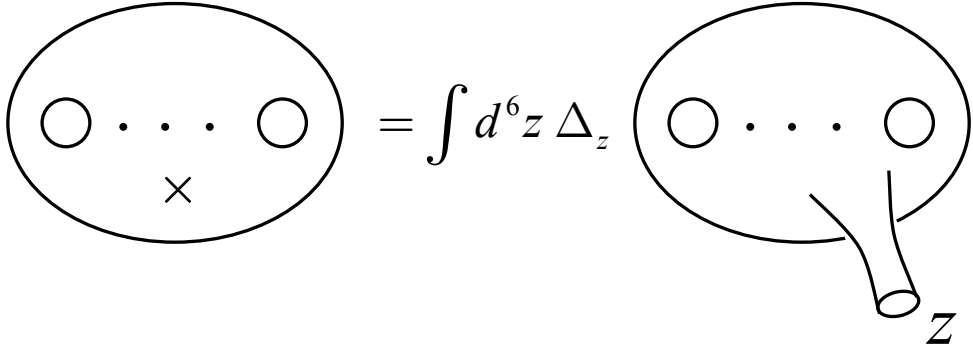}
\caption{LSZ-like reduction. The cross represents the insertion of $\delta S$ or 
$|B_0\rangle$.}
   \label{LSZ-like}
\end{figure}

To evaluate $F_n(|B_0\rangle)$, consider a worldsheet path-integral $F_{n+1}(z)$ with $n+1$ 
boundaries, one of whose boundary is placed at $x^I=z^I$ with $I=5,\cdots,10$. 
Note that the $N-1$ D-instantons we have discussed so far are placed at $x^I=0$. 
In other words, we place another set of D-instantons 
which are distributed parallel to the original D-instantons but their 
positions in $x^I$-directions are different. 
It is possible to obtain $F_n(|B_0\rangle)$ from $F_{n+1}(z)$ through an LSZ-like procedure: 
\begin{equation}
F_n(|B_0\rangle) = \int d^6z\ \Delta_zF_{n+1}(z),  
   \label{LSZ}
\end{equation}
where $\Delta_z$ is the Laplacian on ${\bf R}^6$. 
See figure \ref{LSZ-like} for an image of this procedure. 
The variation of the total free energy is therefore 
\begin{equation}
\delta F(\lambda) = \int d^6z\ \Delta_zF(\lambda,z), 
   \label{totalvariation}
\end{equation}
where 
\begin{equation}
F(\lambda,z) = \sum_{n=0}^\infty \frac{\lambda^n}{n!}F_{n+1}(z). 
\end{equation}

Since a single D-instanton is separated from the hypersurface along which the other D-instantons are 
distributed, the worldsheet is stretched in the region $0\le |x^I|=\sqrt{x^Ix^I}\le r$ where 
\begin{equation}
r = \mbox{max}\{l_s,h\} 
\end{equation}
and $h$ is the distance of 
the separated D-instanton from the hypersurface of the D-instantons. 
If we take $|z^I|$ to be larger than $r$, then the corresponding worldsheet has a thin tube connecting 
the boundary at $x^I=z^I$ and the body of the worldsheet. 
The tube represents the propagation of closed string states, and the dominant contribution comes from 
the massless propagation. 
As a result, $F(\lambda,z)$ behaves as $|z^I|^{-4}$ for large $|z^I|$. 
For the range $0\le |z^I|\le r$, we assume that $F(\lambda,z)$ varies slowly with $|z^I|$. 

Let us calculate the integral 
\begin{equation}
I = \int d^6z\ \Delta_z f(z), 
\end{equation}
where 
\begin{equation}
f(z) = \left\{ 
\begin{array}{cc}
f(0), & (0\le |z^I|\le r) \\ [3mm] \displaystyle{\frac{r^4}{|z^I|^4}f(0)}, & (|z^I|>r)
\end{array}
\right.
\end{equation}
is a typical example of the function having the property assumed for $F(\lambda,z)$. 
It is easy to check that $I=-4\ vol(S^5)r^4f(0)$. 
Similarly, the RHS of (\ref{totalvariation}) would be estimated as 
\begin{equation}
\int d^6z \ \Delta_z F(\lambda,z) \sim r^4C(\lambda,r)F(\lambda,z=0), 
   \label{estimate}
\end{equation}
where $C(\lambda,r)$ is assumed to be of order one. 
Noticing that $F_{n+1}(z=0)=F_{n+1}$, we obtain 
\begin{eqnarray}
\delta F(\lambda) 
&\sim& r^4C(\lambda,r)\partial_\lambda F(\lambda), 
   \label{variation}
\end{eqnarray}
where 
\begin{equation}
\sum_{n=0}^\infty \frac{\lambda^n}{n!}F_{n+1} = \partial_\lambda F(\lambda)
\end{equation}
is used. 
In other words, the free energy transforms as 
\begin{equation}
F(\lambda) \to F(\lambda+\epsilon r^4C(\lambda,r)), 
   \label{result}
\end{equation}
which indicates {\it the existence of the 
scale invariance if $r<<\lambda^{\frac14}$}. 
Note that the sum of the infinite number of worldsheets with 
boundaries is crucial for the existence of this scale 
invariance. 

It should be pointed out that our calculation has been done with an appropriate analytic continuation 
of $\lambda$. 
At first, the free energy is defined as a power series of $\lambda$ as (\ref{free}). 
This definition is valid when the effects from the D-instantons are small. 
However, to estimate the variation of the free energy, 
we assume the behavior of the sum 
$F(\lambda,z)$, not each $F_n(z)$, and therefore, 
this estimate should be valid even 
beyond the convergence radius of the perturbative series (\ref{free}). 
The result that there exists 
a scale invariance if $r<<\lambda^{\frac14}$ agrees with the 
existence of an isometry of the corresponding metric when $\lambda$ is large. 
To see the importance of the analytic continuation, let us consider 
the metric of D-instantons distributed along a four-dimensional hypersurface 
\cite{GGP}
\begin{equation}
ds^2 = H(\rho)^{\frac12}(\eta_{\mu\nu}dx^\mu dx^\nu+d\rho^2+\rho^2d\Omega_5^2), 
   \label{metric} 
\end{equation}
where 
\begin{equation}
H(\rho) = 1+\frac{C\lambda}{\rho^4}. 
\end{equation}
In the region $\rho<<\lambda^{\frac14}$, the metric (\ref{metric}) behaves as 
\begin{equation}
ds^2 \sim \sqrt{C\lambda}\left[ \frac{\eta_{\mu\nu}dx^\mu dx^\nu+d\rho^2}{\rho^2}+d\Omega_5^2 \right], 
\end{equation}
which is a metric on $AdS_5\times S^5$. 
In this metric, the scale invariance is realized as the isometry 
\begin{equation}
\delta x^\mu = x^\mu, \hspace{5mm} \delta \rho = \rho. 
\end{equation}
From the perturbative string point of view, the metric (\ref{metric}) 
is given as a power series of $\lambda$. 
By summing them up, we obtain the non-trivial function $H(\rho)^\frac12$. 
The possibility to have such a closed expression 
enables us to go beyond the convergence radius of the perturbative 
series where the isometry exists. 

\vspace{5mm}

We consider a worldsheet which is relevant to a high energy scattering of open string states on the 
D3-branes. 
As is explained above, 
one of the D3-brane is separated from the other $N-1$ D3-branes to implement the IR 
regularization, and all the external open string states are on the separated D3-brane. 

\begin{figure}[tbp]
\hspace*{5mm}
\includegraphics{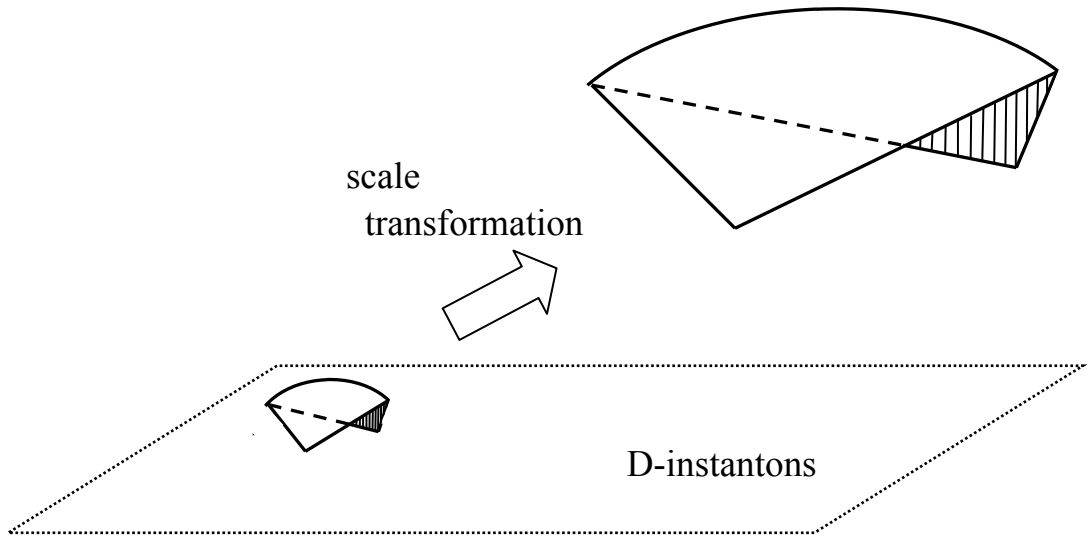}
\caption{Scale transformation in the T-dual picture. }
   \label{scale}
\end{figure}

We make a scale transformation in the T-dual picture which is isotropic and brings the worldsheet 
to a region far away from the D-instantons. 
The situation is depicted in figure \ref{scale}. 
If we take $\lambda$ to be large, then the worldsheet can be brought to a place far enough from the 
hypersurface of the 
D-instantons so that the presence of the D-instantons is represented by the curved background 
(\ref{metric}) 
while the scale invariance is still valid. 
In the region $\rho<<\lambda^{\frac14}$, the background (\ref{metric}) is approximated by 
$AdS_5\times S^5$. 
Then, the classical solution of the string is 
the minimal surface obtained in \cite{AM}. 
The size of the minimal surface is determined by the momenta of the external open string states, 
which become larger and larger by the scale transformation. 
It should be noted that, before the scale transformation, the size of the worldsheet is 
smaller than the string 
scale, as mentioned at the beginning of this section. 
After the scale transformation, 
this classical solution will dominate the summation over worldsheets. 
Note that, 
since the scale transformation for the D-instanton system is isotropic, the ratio of the size of 
the minimal surface to the distance from the hypersurface does not change. 
This implies that the worldsheet always exists near the boundary ($\rho=0$) of $AdS_5$. 
Notice that in the D-instanton case, the ``near horizon'' region corresponds to the region of the 
boundary of $AdS_5$, as can be seen from the metric (\ref{metric}).  
This is in contrast with the situation in \cite{KS} where D3-branes are placed at the center of the 
$AdS_5$ space-time from the gravity point of view. 

Since the scale transformation is a symmetry as long as the worldsheet is restricted in the region 
$|x^I|<<\lambda^{\frac14}$, the classical action for the minimal surface provides the 
high energy scattering 
amplitude with which we started. 
In this way, the scale invariance found in \cite{KS} provides the argument which verifies the 
relation 
claimed in \cite{AM}. 

One may think that the approximate symmetry is broken by the presence of the non-trivial dilaton 
background 
\begin{equation}
e^{\Phi} = H(\rho). 
\end{equation}
Recall that the dilaton coupling in the worldsheet action is a sub-leading order term of $\alpha'$. 
Since the actual expansion parameter is $\frac{\alpha'}{R^2}$ 
where $R$ is a typical length scale of the 
target space-time, which is proportional to $\lambda^{\frac14}$ in this case, 
the contribution from the dilaton background would be negligible if we take 
a large $\lambda$.

\vspace{5mm}

\vspace{1cm}

\section{Finite temperature}

\vspace{5mm}

Next, we consider D3-branes at finite temperature. 
This is realized by considering a Euclideanized D3-brane which wraps on a circle with the 
circumference $\beta$. 
$AdS$/CFT correspondence has also been considered in this case \cite{Witten}. 
The argument on the scale transformation can be also carried out in the case of the finite temperature. 

One of the crucial points of our perturbative approach 
is the transformation property of the boundary state of the wrapped 
D3-brane. 
Since we impose the anti-periodic boundary condition for space-time fermions in the Euclidean time 
direction on which the D3-brane wraps, the fermionic worldsheet variables flip their signs as the 
string winds around the direction. 
The scale transformation for these winding sectors is the same as that for the zero-winding sector, 
except for the fact that the modings of $S^a$ and $\tilde{S}^a$ are half-integral if the winding 
number is an odd integer. 
Due to the presence of these winding sectors, the boundary state has the form 
\begin{equation}
|B\rangle = \sum_{w\in{\bf Z}}|B;w\rangle, 
\end{equation}
where $w$ is the winding number, and 
\begin{eqnarray}
|B;2k\rangle &=& \exp\left[ \sum_{n=1}^\infty\left( \frac1nM^{ij}\alpha^i_{-n}\tilde{\alpha}^j_{-n}
 -iM^{ab}S^a_{-n}\tilde{S}^b_{-n} \right) \right]|B_0;2k\rangle, \\
|B;2k+1\rangle &=& \exp\left[ \sum_{n=1}^\infty\left( \frac1nM^{ij}\alpha^i_{-n}\tilde{\alpha}^j_{-n}
 -iM^{ab}S^a_{-n+\frac12}\tilde{S}^b_{-n+\frac12} \right) \right]|B_0;2k+1\rangle. 
\end{eqnarray}
The term $|B;2k\rangle$ has almost the same form with the supersymmetric boundary state \cite{GG}, and 
the transformation property is the same except for the obvious scaling of the circumference $\beta$. 
It might look 
non-trivial to analyze the transformation property of the term $|B;2k+1\rangle$, due to the 
half-integral moding of the fermionic oscillators. 
However, one can check that the calculations can be done similarly with the case of $|B;2k\rangle$, 
and obtain 
\begin{equation}
\delta |B;2k+1\rangle \propto |B;2k+1\rangle.
\end{equation}
Therefore, we can 
define the scale transformation so that the boundary state $|B\rangle$ is scale invariant up to 
the scaling of the circumference $\beta$. 
Then, $AdS$/CFT correspondence also follows at finite temperature case, as in \cite{KS}. 
Note that the length scale of the background, if exists, is also transformed by the scale 
transformation. 
This fact implies that, if one considers a Wilson loop realized as in \cite{KS}, then it is related to 
a minimal surface placed at the outside of the event horizon of a black hole background, since the 
boundary of the worldsheet is always at the outside of the event horizon. 

\vspace{5mm}

There are other researches in the case at finite temperature 
which discuss the entropy of ${\cal N}=4$ 
SYM and that of the $AdS$-Schwarzschild black hole \cite{GKP}\cite{Witten}. 
Due to the limitation of the explicit calculations, the entropy is calculated in the classical 
gravity only when $\lambda$ is large, and in the SYM only when $\lambda$ is small. 
The corrections to these results have also been calculated in 
\cite{FT}\cite{M}\cite{KR}\cite{GKT} 
which suggest that the entropy in the 
gravity region is smoothly interpolated to the entropy in the SYM region by varying $\lambda$. 
These are the arguments supporting that the SYM entropy and the black hole entropy are the same 
for any $\lambda$. 
In the following, we will show the coincidence of the entropy at large $\lambda$. 

Our point of view on $AdS$/CFT correspondence is based on D3-branes in the flat space-time. 
The temperature is encoded in the radius of the Euclideanized time direction. 
The temperature-dependence thus comes from strings which wind around the time direction. 
Let us consider the free energy of strings in this case. 
In the large $N$ limit with $\lambda$ kept fixed, the winding closed strings with the genus $h\ge1$ 
provide contributions of order $g_s^{2h-2}\sim\lambda^{2h-2}N^{2-2h}$ 
which is at most of order $N^0$. 
On the other hand, the winding open strings may provide contributions of order $N^2$. 
Therefore, in the large $N$ limit, the temperature-dependent part of the 
free energy is dominated by open strings. 
When the temperature is small, then only massless open string states contribute. 
As a result, the thermodynamical quantities which are obtained from the temperature-dependence of the 
free energy, for example the energy and the entropy, are equal to those of the SYM in the limiting case 
mentioned above. 

From an observer at the asymptotically flat region, the thermal D3-branes are regarded as a 
non-extremal black hole 
\begin{equation}
ds^2 = H(r)^{-\frac12}(-f(r)dt^2+d\vec{x}^2)+H(r)^{\frac12}(f(r)^{-1}dr^2+r^2d\Omega_5^2), 
   \label{nonext}
\end{equation}
where 
\begin{equation}
H(r) = 1+\frac{4\pi\lambda}{r^4}, \hspace{5mm} f(r) = 1-\frac{r_0^4}{r^4}. 
\end{equation}
The constant $r_0$ is determined so that the Euclideanized version of (\ref{nonext}) does not have 
a conical singularity. 
If $\lambda$ is taken to be large, then $r_0$ can be written as 
\begin{equation}
r_0 = \frac{\pi\lambda^{\frac12}}{\beta}. 
\end{equation}
Note that $\beta$ here is the circumference of the Euclidean time circle at the asymptotically flat 
region. 
Therefore, this $\beta$ coincides with the one appeared in the D-brane setup. 
In this way, we can equate the temperature of the gravity side with that of the gauge theory side. 

The gravity description is valid when $r_0$ is large. 
If we take $\lambda$ to be large, we can also take $\beta$ 
to be large while keeping $r_0$ still large. 
This means that there exists a parameter region in which both the gravity description and the SYM 
description are valid. 
(The description in terms of free SYM cannot be valid in this region, 
of course.) 
The ADM mass of the black hole must be the same with the total energy of the SYM 
plus the contribution from the tension 
of the D3-branes. 
Since the entropy is obtained from the thermodynamic relation
\begin{equation}
\frac{\partial S}{\partial E} = \frac1T, 
\end{equation}
we can conclude that 
the entropies of the SYM 
and the black hole must be the same, due to the coincidence of the energy and the temperature, 
although it is difficult to perform an explicit evaluation of 
the entropy of the SYM with large $\lambda$. 

The crucial points of our argument are that the temperature-dependence of the free energy is dominated 
by open strings in the large $N$ limit, and that there is a region of parameters where both the 
gravity description and the SYM description are valid. 
It is also important that some of physical quantities, the energy and the temperature for example, 
in two descriptions can be compared directly 
with each other at the asymptotically flat region which is absent after taking the near-horizon 
limit.

\vspace{1cm}

\section{Discussion}

\vspace{5mm}

We have shown that our point of view of $AdS$/CFT correspondence, based on the perturbative Type IIB 
string 
theory with D3-branes, can be applied to some situations discussed in the literature of $AdS$/CFT 
correspondence. 
The amplitude of high energy scattering processes of the SYM can be related to classical worldsheet 
configurations in the $AdS_5$ space obtained in \cite{AM} by the scale transformation, which clarifies 
the reason why such a correspondence holds. 
Thermal properties of the SYM are also discussed in our point of view. 
It can be shown that the reduction of open string system on D3-branes to the SYM occurs even in the 
case of large $\lambda$, and some thermal properties of the SYM can be related to some gravitational 
quantities, the connection being made at the asymptotically flat region far from the D3-branes. 

The successful applications of our point of view indicate that it would enable us to make further 
predictions for the correspondence which are not known so far. 
As long as the correspondence is based on our scale invariance, the coincidence of some quantities 
will have a firm footing. 
We hope to provide some new and non-trivial correspondences between gauge theory and gravity, possibly 
less supersymmetric, which could be checked by explicit calculations.

\vspace{2cm}

{\bf \large Acknowledgments}

\vspace{5mm}

We would like to thank M.Fukuma, T.Matsuo, T.Takayanagi, T.Uematsu for valuable discussions. 
We would also like to thank L.Dixon for valuable comments on the IR regularization of the SYM 
amplitudes. 
This work is supported by the Grant-
in-Aid for the 21st Century COE "Center 
for Diversity and Universality in Physics" from the Ministry of Education, 
Culture, Sports, Science and Technology (MEXT) of Japan.
The research of T.S is supported in part by JSPS Research Fellowships for Young Scientists.

\end{document}